\documentclass[
  ,final            
  ]
  {aipproc}
  
\layoutstyle{6x9}

\begin{document}

\title{Kaluza--Klein Dark Matter from Deconstructed Universal Extra
Dimensions\footnote{Talk presented at SUSY06, the 14th International
Conference on Supersymmetry and the Unification of Fundamental
Interactions, UC Irvine, California, 12-17 June 2006.}}

\classification{11.10.Kk,12.60.-i,11.15.Ha}
\keywords      {Dark Matter, Beyond Standard Model, Field Theories in Higher Dimensions}

\author{Tomas H\"allgren\footnote{E-mail adress: tomashal@kth.se}}{
  address={Department of Theoretical Physics, School of Engineering
Sciences, Royal Institute of Technology (KTH), AlbaNova University
Center, Roslagstullsbacken 21, 106 91 Stockholm, Sweden}}

\begin{abstract}
We consider Kaluza--Klein dark matter from deconstructed or latticized
universal extra dimensions and study in this model the positron flux
from Kaluza--Klein dark matter annihilation in the galactic halo.


\end{abstract}

\maketitle


\section{Introduction}

This talk is based on Ref.~\cite{Hallgren}. Today there is strong
cosmological evidence for the presence of non-luminous dark matter
\cite{Spergel}. Measurements indicate that the
energy and matter in the Universe should be distributed such that
approximately 73~\% is ``dark energy'', 23~\% is ``dark matter'', and
around 4~\% is ordinary luminous matter. In recent years, an
interesting alternative WIMP dark matter candidate, Kaluza--Klein dark
matter (KKDM), has been intensively studied \cite{Servant,Cheng}. The
WIMP candidate is in this case the lightest Kaluza--Klein particle
(LKP) in models with universal extra dimensions (UEDs), usually taken
to be the first excited mode of the hypercharge gauge boson. It is
stable due to the conservation of Kaluza--Klein (KK) parity and the
detection prospects have been shown to be good \cite{Cheng}. Due to
the close degeneracy of the KK spectrum, radiative corrections to KK
masses could be crucial when determining the nature of the
LKP. However, higher dimensional field theories are not renormalizable
and unknown contributions from the UV-completion could therefore be
essential. Recently, a possible UV-completion of higher dimensional
gauge theories, known as deconstructed or latticized extra dimensions
\cite{Arkani-Hamed}, was suggested\footnote{Usually, gravity is not included. For recent work on
discretized gravity in 6D warped space, see \cite{Bauer}.}. In these
type of models, the higher-dimensional theory is interpreted as a
non-linear sigma model, which can be completed in the ultraviolet, by
for example a linear sigma model in 3+1 dimensions. In such a setting,
one could calculate radiative corrections to KK masses
\cite{Falkowski}. We will address the possibility of KKDM 
in deconstructed universal extra dimensions \footnote{We will not
discuss radiative corrections.}. We will follow
\cite{Wang,Oliver}, in particular \cite{Oliver}, here extended
to the leptonic sector. We will also discuss the prospects for
indirect detection.

\section{The Model}

The fundamental model which we will consider is a linear sigma model
in 3+1 dimensions with a product gauge group
$G=\Pi_{j=0}^{N}SU(3)_{j}\times SU(2)_{j} \times U(1)_{j}$. The model
contains fermions, gauge bosons as well as a set of link fields,
$Q_{j,j+1},\Phi_{j,j+1}$, and $\phi_{j,j+1}$, which transform in the
bifundamental representation of adjacent $SU(3)$, $SU(2)$, and $U(1)$
gauge groups, respectively. When the link fields acquire vacuum
expectation values, {\it i.e.},$\langle
Q_{j,j+1}\rangle=v_{3}\bf{1_{3}}$, $\langle
\Phi_{j,j+1}\rangle=v_{2}\bf{1_{2}}$, and $\langle
\phi_{j,j+1}\rangle=v_{1}/\sqrt{2}$, the product gauge group is broken
down to the diagonal subgroup, which we identify as the standard model
(SM) gauge group. At the same time, the kinetic terms for the link
fields generates mass matrices for the gauge bosons, which after
diagonalization yield for the $U(1)$ gauge bosons the
spectrum
\begin{equation}\label{eq:gbmass}
m_{n}^{2} = \tilde{g}_{Y}^{2} v_{1}^{2} Y_{\phi}^{2} \sin^{2}
\left[\frac{n\pi}{2(N+1)}\right]
\end{equation}
and similarly for the $SU(2)$ and $SU(3)$ gauge bosons. This becomes,
with the identification $\pi \tilde{g}_{Y} v_{1} Y_{\phi}
/[2(N+1)]=1/R$ and in the limit $n\ll N$ indistinguishable from a
usual linear KK spectrum $m_{n} \simeq n/R$. We identify the first
excited mode $\tilde{A}_{1}$ as the LKP and dark matter candidate.

We include a set of fermions $L^{\alpha}_{j}=(\nu^{\alpha}_{j},
e^{\alpha}_{j})^{T}$ and $E^{\alpha}_{j}$, for $j=0,1,\ldots,N$ and
$\alpha=e,\mu,\tau$ (from now on flavor indices will be
suppressed). Here $L_{j}$ transforms as ${\bf 2}$ under $SU(2)_{j}$
and as a singlet under $SU(2)_{i}$, for $i\neq j$. Furthermore,
$L_{j}$ is charged under $U(1)_{j}$ as $Y_{d}=-1$. The field $E_{j}$
is a singlet under all $SU(2)$ groups and is charged under $U(1)_{j}$
as $Y_{s}=-2$. Both $L_{j}$ and $E_{j}$ transform trivially under all
$SU(3)$ groups.  In Ref.~\cite{Oliver}, the latticized action for the
quark sector was considered. We construct analogously the action for
leptons as $S_{{\rm fermion}}=S_{d}+S_{s}$, where $S_{d}$ refers to
the part containing the $SU(2)$ doublet fields and $S_{s}$ contains
the $SU(2)$ singlet fields. The action $S_{d}$ is obtained from a
naive discretization of the continuum action augmented by a Wilson
term, and is given by
\[
S_{d} = \int {\rm d}^{4}x\left\{\sum_{j=0}^{N}\bar{L}_{j}{\rm
i}\gamma^{\mu}D_{\mu}L_{j} - \sum_{j=0}^{N} \left[ M_{f}{\bar L}_{jL}
\left( \frac{\Phi^{\dagger}_{j,j+1}}{v_{2}}
\frac{\phi_{j,j+1}^{3}}{(v_{1}/\sqrt{2})^{3}} L_{j+1,R} - L_{jR}
\right) + {\rm h.c.} \right]\right\}, \nonumber
\]
where $M_{f}$ is a mass parameter which is used when matching to the
continuum model. The expression for $S_{s}$ is obtained similarly.

In order to obtain chiral zeroth modes, we take the doublet fields to
satisfy $L_{0R}=0$ and the singlet fields to satisfy $E_{0L}=0$. When
the link fields acquire universal VEVs, we obtain mass matrices for
the fermion fields, which after diagonalization give for the
left-handed doublet fields masses of the form as in
Eq.~(\ref{eq:gbmass}), with $\tilde{g}_{Y}^{2} v_{1}^{2} Y_{\phi}^{2}
\rightarrow 4 M_f^{2}$, where $n=0,1,\ldots, N$ and for the
right-handed doublets fields masses of the same form, but now
$n=1,2,\ldots, N$. Thus, there are no zeroth modes for the
right-handed doublet fields. For the singlet fields the reverse
situation holds. This reproduce the feature of chirality of the zero
modes, as in the continuum theory. For $n\ll N$, we find a linear KK
spectrum if we make the identification $\pi M_{f}/(N+1)=1/R$. It is
straightforward to include also electroweak symmetry breaking
masses. The Feynman rules for the fermion and gauge boson interactions
can be read off from the fermionic kinetic terms and the gauge boson
field tensor terms, using the mode expansions. For the $L_{jL}$ and
$E_{jR}$ interactions, as well as for the gauge boson
self-interactions, KK-parity is conserved, even for a model with only
few sites, which means that there are no decay channels in this
sector. For the $L_{jR}$ and $E_{jL}$ sectors we find that for a
few-site model KK parity is only approximately conserved in certain
decay channels of the $\tilde{A}_{1}$ mode (the LKP) at loop
level. These decay channels were not taken into account in
\cite{Hallgren}. However, for large $N$, KK parity becomes an
arbitrarily good symmetry, also in these sectors of the model. The
analysis for the quark sector is analogous. What is also needed in
order to have the LKP as the dark matter candidate, are additional KK
parity conserving interactions which violate KK number. These are in
the continuum formulation generated at one-loop level from the
orbifold compactification \cite{Schmaltz}. A similar analysis could be
applied also to the deconstructed model, where deconstruction, in the
sense of a UV-completion, could be useful when addressing
contributions from UV-physics, which in the usual continuum model are
not calculable. Here we will assume that such KK number violating
interactions are present and leave the detailed analysis for future
work.

\section{Indirect Detection}

In this section, we consider the positron flux from KKDM ({\it i.e.}
$\tilde{A}_{1}$) annihilation in the galactic halo, for the lattice
model described in the previous section. Here, as in \cite{Cheng} we
only consider positrons from direct $e^{+}e^{-}$ production. The
analysis will be similar to that of the continuum case, with the
linear continuum KK mode spectrum replaced by the non-linear spectrum
of the lattice model.

\begin{figure}
\includegraphics[height=.3\textheight]{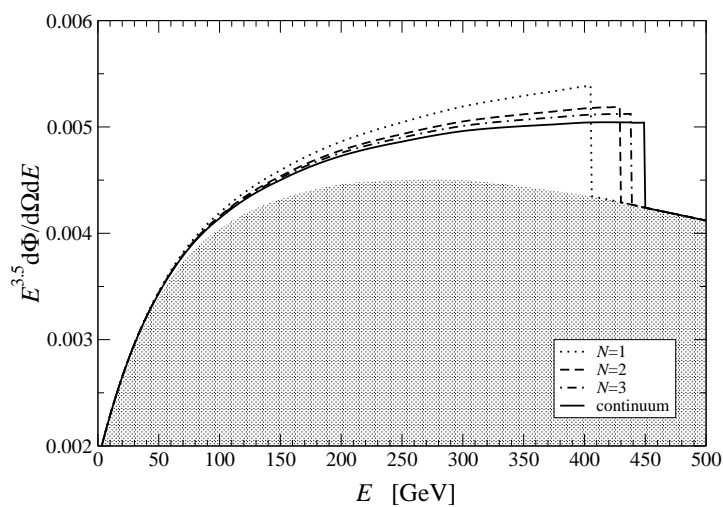}\caption{{\small The
differential positron flux (above background) for an inverse radius of
450~GeV as a function of positron energy, for direct $e^{+}e^{-}$
production. Presented are latticized models with two lattice sites
($N=1$, dotted curve), three lattice sites ($N=2$, dashed curve) and
four lattice sites ($N=3$, dash-dotted curve) as well as the continuum
model (solid curve). Given is also an estimated background flux (gray
shaded). The unit of the ordinate is
cm$^{-2}$s$^{-1}$sr$^{-1}$GeV$^{2.5}$. Figure from
\cite{Hallgren}.}}\label{fig:plot}
\end{figure}

In calculating the differential positron flux, we have used the {\sc
DarkSUSY} package, see \cite{Gondolo}. If the model in \cite{Hallgren}
is extended to include the KK parity violating one-loop decay channels
mentioned earlier, then for a model with only few sites there may not
be a viable dark matter candidate. A quantitative study is left for
future work. For a large number of sites, KK parity becomes an
arbitrarily good symmetry and the dark matter candidate is stable. In
Fig.\ref{fig:plot}, we present the differential positron flux as a
function of the positron energy for an inverse radius of 450~GeV, for
lattice models with $N=1$ ({\it i.e.}, two lattice sites), $N=2$,
$N=3$, and the continuum model results. The parameter values are from
Ref.~\cite{Hallgren}.

The bounds on the mass of the first excited KK mode, coming from, for
example, electroweak precision tests (EWPT), limits the prospects for
indirect detection, with for example the PAMELA
\cite{pamela} and AMS-02 \cite{AMS-02} experiments. It was shown in
Ref~\cite{Oliver} that the EWPT bounds are lowered for a lattice
model, by as much as 10~\%-25~\%, for a few-site lattice model, the
reason being the realization of a finite number of KK
modes. Therefore, a lattice model could, in principle, improve the
detection prospects for PAMELA and AMS-02.

The peak in the positron spectrum is as in the continuum model due to
the monoenergetic positron source and is a characteristic signature of
KKDM, distinct from the signal from neutralinos.

\begin{theacknowledgments}
I would like to thank the organizers of SUSY06 for a very nice
conference and Tommy Ohlsson for the collaboration on this project. I
would also like to thank the G{\"o}ran Gustafsson Foundation and the
Styffes Foundation for financial support.
\end{theacknowledgments}

\end{document}